\documentstyle[12pt]{article}

\def\sp#1{{}^{#1}}                              
\def\sb#1{{}_{#1}}                              

\def\a{\alpha}
\def\b{\beta}
\def\g{\gamma}
\def\d{\delta}
\def\c{\chi}
\def\f{\phi}
\def\e{\epsilon}
\def\th{\theta}
\def\l{\lambda}
\def\D{\Delta}
\def\S{\Sigma}
\def\pa{\partial}

\newcommand{\be}{\begin{equation}}
\newcommand{\ee}{\end{equation}} 
\newcommand{\bea}{\begin{eqnarray}}
\newcommand{\eea}{\end{eqnarray}}

\begin{document}

\begin{titlepage}

\begin{flushright} 
{\tt 	 FTUV/97-65\\IFIC/97-95\\UALG/TP/97-2}
 \end{flushright}

\bigskip

\begin{center}

{\bf{\LARGE Remarks on the Reduced Phase Space of (2+1)-Dimensional Gravity 
on a Torus
in the Ashtekar Formulation}}

\bigskip 
A.  Mikovi\'c$^{1}$\footnote{ E-mail address:
 mikovic@lie.ific.uv.es} and
N. Manojlovi\'c$^{2}$\footnote{E-mail address: nmanoj@ualg.pt}

\end{center}


\footnotesize

\begin{enumerate}	                 
\item Departamento de F\'{\i}sica Te\'orica and 
	IFIC, Centro Mixto Universidad de Valencia-CSIC,
	Universidad de Valencia,	
        Burjassot-46100, Valencia, Spain and Institute of Physics,
P.O. Box 57, 11001 Belgrade, Yugoslavia
\item  Unidade de Ciencias Exactas e Humanas, Sector de Matematica, 
Universidade do Algarve, Campus de Gambelas, 8000 Faro, Portugal.
\end{enumerate}
\normalsize 

\bigskip

\begin{abstract}			

We examine the reduced phase space of the
Barbero-Varadarajan solutions of the Ashtekar formulation of
(2+1)-dimensional general relativity on a torus.  
We show that it is a finite-dimensional space due to existence of an infinite
dimensional residual gauge invariance which reduces the infinite-dimensional
space of solutions to a finite-dimensional space of gauge-inequivalent 
solutions. This is in agreement with general arguments which imply that
the number of physical degrees of freedom for (2+1)-dimensional
Ashtekar gravity on a torus is finite.

\end{abstract}



\end{titlepage}

\newpage

\section{Introduction}

Ashtekar formlation of General Relativity (GR) \cite{asht}
takes the spatial connection as the basic dynamical variable,
and this has been very fruitfull idea, especially for the formulation of
a consistent quantum theory of GR. The Ashtekar formulation can be also 
applied to (2+1)-dimensional GR \cite{beng}. 
(2+1)-dimensional GR is a very useful toy model of
quantum gravity \cite{carlip},
since it boils down to quantum mechanics while
it has enough structure so that various conceptual problems of quantum gravity
can be examined. The main reason for this is that the 
reduced phase space (rps) for (2+1)-dimensional GR in the
metric formulation is of finite dimension. This also happens in the Witten
formulation of (2+1)-dimensional GR, which is an alternative connection 
formulation \cite{wit}. 

Since in the
connection formulations the metric is a derived quantity, the phase space
contains points corresponding to the degenerate spatial metric. One can show 
that the Ashtekar formulation is equivalent to the Witten formulation for
non-degenerate metrics \cite{beng}, but in the degenerate sector they are
non-equivalent \cite{mm}. On the basis of general arguments, one expects that
the total reduced phase space of the Ashtekar formulation is also a finite
dimensional symplectic manifold, or a finite union of the former
(see \cite{mm} for the case of toroidal spatial section). However,
Barbero and Varadarajan (BV) have made a claim that the rps for torus is 
infinite-dimensional \cite{bv}. 
This is done by considering special connection configurations which contain
the degenerate metric sectors, and then solving the constraints.
In this way one obtains
solutions which depend on arbitrary many parameters. Since these parameters 
are integrals of motion, one concludes
that the total rps is infinite-dimensional. This is in contrast to the
result of \cite{mm}, where it was argued that the total rps is of finite 
dimension.

In this paper we show that every BV solution can be related by a gauge 
transformation to the solution with a constant connection. This means that
BV parameters are not gauge invariant objects or observables, although they
are integrals of motion. This follows from the fact that BV parameters are
Wilson loops, which are not observables in the Ashtekar formulation, although
they can be integrals of motion in gauges where the connection is flat.
This is also reflected by the fact that the BV gauge for the connection
has an infinite-dimensional 
residual gauge invariance which corresponds to the fact that the gauge
is used where the constraints become linearly
dependent. Consequently, one can gauge-fix further, so that  the spatial
coordinate dependence of the corresponding degrees of freedom (dof) is 
removed. Hence one 
is left with only one gauge inequivalent dof.  

In section two we give a general argument why  the rps for (2+1)-dimensional
Astekar gravity on a torus is of finite dimension, while in section three
we explicitely demonstrate our claims about the BV solutions. In section four
we present our conclussions.  

\section{General considerations}

It is well known that for a gauge theory with $m$ independent gauge 
symmetries, the number of physical dof is given by $n-m$ fields per 
spatial point plus 
$n-m$ corresponding canonicaly conjugate momenta, where $n$ is the number of 
dynamical components of the gauge field. In the case when $m = n$,
there are no local dof, since one can choose a gauge where  $q\sp i (x)$
are independent of the spatial coordinates $x$, or more generally,
depend on a finite number of constant parameters. The formal
argument for this statement is the following. Let $G_i (p,q)$ be the  
first-class
irreducible constraints on the phase space $(p_i(x) , q\sp i (x))$, 
$i=1,...,n$. Solve the constraints $G_i =0$ for the momenta $p_i$ as
$p_i = f_i (q, C_\a)$, where $C_\a$ are constants of integration, and define
\be P_i (x)  = p_i (x) - f_i (q(x), C_\a ) \quad.\label{11}
\ee
The new variables $P_i$ satisfy $\{ P_i , P_j \} =0$ \cite{henn,kuch}, and 
$P_i =0$ is equivalent to $G_i =0$. We can consider $P_i$ as the new momenta,
and therefore introduce $Q\sp i$ as the corresponding coordinates \cite{fadd}.
Therefore the canonical transformation $(p,q) \to (P,Q)$ makes the constraints
abelian. The corresponding gauge invariance also becomes abelian, and it
is given by
\be \d P_i (x) = 0 \quad,\quad \d Q\sp i (x) = \e\sp i (x) \quad. \label{12}\ee

By using (\ref{12}) one can set each $Q\sp i$ to zero, and inequivalent
solutions will be labeled by $C_\a$. These are the rps coordinates, and for
reparametrization invariant theories $C_\a$ are observables and hence
integrals of motion. What is
less clear from (\ref{11}), is that there are finitely many $C_\a$. Proving
this in the general case may be difficult, but for a concrete theory, one can
demonstrate this by showing that a gauge
\be q\sp i (x) = F\sp i (Q_I, x) \quad,\label{13}\ee
where $\{ Q_I \} $ is a finite set of global ($x$-independent) coordinates,
can be allways choosen by performing a gauge transformation  on an arbitrary 
configuration $q\sp i (x)$. In the case of (2+1)-dimensional
Ashtekar gravity on a torus, this 
was demonstrated in \cite{mm}, although not explicitely, 
so that the dynamics boils down to a constrained particle system. 

The explicit argument behind the ansatz from \cite{mm} is the following.
The Ashtekar constraints for (2+1)-dimensional GR can be written as
\bea
G\sb a &=& D_i E\sp{i}\sb a = \pa_i E\sp i\sb a + \e_{abc}A_i\sp b E\sp{ic}
\quad,\label{14}\\
G_i   &=& F_{ij}\sp a E\sp i\sb a \quad,\label{15}\\
G_0  &=&  F_{ij}\sp a E\sp{ib} E\sp{jc}\e_{abc} \quad,\label{16}\eea
where $A$ is an $SO(1,2)$ connection one-form on a spatial section $\S$, 
$F$ is the corresponding curvature two-form 
($F_a = d A_a + \e_a\sp{bc} A_b \wedge A_c$) 
and $E$ is the canonicaly conjugate
vector density, such that $g\sp{ij} = E\sp{ia} E\sp j\sb a$ is a metric density
on $\S$. In the case when $\Sigma$ is a torus, one can introduce
new phase space variables $({\cal A}\sb\a\sp a (x) ,{\cal E}\sp\a\sb a(x))$,
$\a =1,2$, via
\be A_i\sp a (x) = {\cal A}\sb\a\sp a (x) \c\sp \a\sb i (x)\quad,\quad 
E\sp i\sb a (x) = {\cal E}_a\sp\a (x) L\sb\a\sp i (x) \quad, 
\label{17}\ee
where $\c\sp\a$ are globally defined one-forms, and $L_\a$ are the 
corresponding vector fields, which satisfy
\be d\c\sp\a = 0 \quad,\quad \c_i\sp\a L\sp i\sb\b = \d_\a\sp\b \quad,\quad 
\oint_{\g_\a} \c\sp\b = \d_\a\sp\b \quad,\label{18} 
\ee
where $\g_\a$ are the homology basis curves. The new variables are
adapted to the global geometry of the torus. By inserting (\ref{17})
into (\ref{14}-\ref{16}) one obtains an 
equivalent system of six constraints for six configuration variables per 
space point. Then via (\ref{11}) and (\ref{12}), one can set each 
${\cal A} (x)$ to a constant.

Alternatively, one can show that there are finitely many $C_\a$ by assuming
the opposite, which would mean that there is a local physical dof, since
in that case one could construct $C(y)= \sum_\a C_\a u_\a (y)$, where 
$u_\a (y)$ is a basis for functions on a sub-manifold of $\S$. This ammounts to
introducing a coordinate dependence in $Q_I$ from (\ref{13}), which is
the case for the BV solutions.
In that
case one should show that there is an infinite-dimensional local residual
gauge invariance for the gauge choice (\ref{13}), which reduces the number
of dof to a finite number. Both approaches will be employed in the case of the
BV solution.

\section{BV solution}

Consider the
following one-Killing vector reduction ansatz of the Ashtekar phase-space
variables \cite{bv} 
\bea 
E\sp\th\sb a &=& E_1 (\th) x_a \quad,\quad A\sb\th\sp a = A_1 (\th) x\sp a 
\label{201}\\
E\sp\f\sb a &=& E_2 (\th) y_a + E_3 (\th )t_a \quad,\quad 
A\sb\f\sp a = A_2 (\th) y\sp a + A_3 (\th )t\sp a \label{202}\eea
where $(\th ,\f)$ are torus coordinates, and $x$ and $y$ are spacelike
vectors, while $t$ is a timelike vector,  forming a basis
in the Lie algebra $so(1,2) \approx sl(2,R)$. We take a basis $J_a$, 
$a =1,2,3$, in the fundamental representation of $sl_2$ algebra
so that $[J_a , J_b ] = \e_{ab}\sp c J_c $ where $\e_{123} = 1$ and
$\eta_{ab} = 2 tr (J_a J_b) = diag (1,1, -1)$. Therefore 
$x = J_1$, $y = J_2$ and $t = J_3$. By
inserting the ansatz (\ref{201}-\ref{202}) into (\ref{14}-\ref{16}) 
we obtain the following constraints
\bea 
G_1 &=& E_1\sp{\prime} + A_2 E_3 - A_3 E_2 = 0, \label{21} \\
G_2 &=& E_2 f_2 - E_3 f_3 =0, \label{22}\\
G_3 &=& E_1 ( E_2 f_3 - E_3 f_2 ) = 0, \label{23}
\eea
where $f_2 = A_2\sp{\prime} - A_1 A_3$ and $f_3 = A_3\sp{\prime} - A_1 A_2$ 
are the non-zero components of $F$ and
$\prime = d/d\th$. The corresponding dynamical system is defined by the
action
\be S = \int dt \int_0^{2\pi} d\th \left( E_i \dot{A}_i - \l\sp i G_i \right)
\quad,\label{231} \ee
where $\l\sp i$ are the Lagrange multipliers enforcing the constraints
(\ref{21}-\ref{23}). The constraints generate the gauge invariance of the
action
\be \d A_i = \int_0^{2\pi} d\th \{ \e\sp j G_j , A_i \}\quad
,\quad \d E_i = \int_0^{2\pi} d\th \{ \e\sp j G_j , E_i \} \ee
 \be \d\l\sp i = \dot{\e\sp i} +\int_0^{2\pi}
 \int_0^{2\pi} d\th_1 d\th_2 \l\sp j (\th_1) \e\sp k 
(\th_2) f_{jk}\sp i (\th_1 , \th_2, \th) 
\quad,\label{232}\ee
where $f$ are the structutre functions of the constraint algebra.
The explicit gauge transformations for the phase-space variables are given by
\bea
\d A_1 &=& -{d\e\sp 1 \over d\th} + (E_2 f_2 - E_3 f_3 )\e\sp 3
\label{233}\\
\d A_2 &=& -A_3\e\sp 1 + f_2 \e\sp 2 + E_1 f_3 \e\sp 3  \label{234}\\
\d A_3 &=& A_2 \e\sp 1 - f_3 \e\sp 2  - E_1 f_2 \e\sp 3   \label{235}
\eea
and
\bea
\d E_1 &=& (E_2 A_3 - E_3 A_2 )\e\sp 2 +  (E_2 A_2 - E_3 A_3 )E_1 \e\sp 3   
\label{236}\\
\d E_2 &=& -E_3\e\sp 1 + (E_2 \e\sp 2 )\sp{\prime} - A_1 E_3 \e\sp 2
- (E_1 E_3 \e\sp 3 )\sp{\prime} + A_1 E_1 E_2 \e\sp 3   \label{237}\\
\d E_3 &=& E_2\e\sp 1 - (E_3 \e\sp 2 )\sp{\prime} + A_1 E_2 \e\sp 2
+ (E_1 E_2 \e\sp 3 )\sp{\prime} - A_1 E_1 E_3 \e\sp 3 \quad. \label{238}
\eea
The equations of motion can be obtained from (\ref{233}-\ref{238}) by replacing
the variations with the time derivatives and the gauge parametars with 
the lagrange multipliers.
 
An immediate indication that the dynamical system 
(\ref{231}) will not 
posses local dof is the fact that there are 3  functionally
independent constraints
acting on 3 configuration variables per space point. This can be demonstrated
in the following way. Let us choose a gauge for the connection such that $F$
is non-null, i.e. $f_2\sp 2 - f_3\sp 2 \ne 0$. Then the constraints imply
\be
E_1\sp{\prime} = E_2 = E_3 =0 \quad.\label{241}
\ee
Now let us choose a gauge where $F$ is null, i.e. $f_2 = \pm f_3$. Then
the constarints give
\be
E_1\sp{\prime} = (A_3 \mp A_2 )E_2 \quad,\quad E_2 = \pm E_3  \quad,\label{242}
\ee
where $E_2$ is arbitrary. It looks like the configurations (\ref{242})
have a local dof. However, by performing a gauge transformation on a 
null connection, one can always reach a non-null connection, since
\be \d (f_2 \mp f_3) = [ 2\e -(A_3 \pm A_2)\e\sp 1 ]\sp{\prime} \pm
A_1 [ 2\e -(A_3 \pm A_2)\e\sp 1 ] \pm 
(A_3 - A_2)d\e\sp 1 /d\th \, ,\label{243}\ee
where $\e =\pm f_2(\e\sp 2 \pm E_1 \e\sp 3 ) $, so that one can always choose
the $\e$'s such that the left hand side becomes non-zero.
Therefore the configuration (\ref{242}) is gauge equivalent to (\ref{241}).
The same applies to the $f_2 = f_3 =0$ configuration. Hence the gauge 
inequivalent solutions are labeled by (\ref{241}), which in turn are labeled
by $E_1 = e_1 =$ const. and $A_1 = a_1=$ const. canonical pair. This 
corresponds to the fact that
\be a_1 = \int_0^{2\pi} d\theta A_1 \label{244}\ee
is the true integral of motion, or the observable, since its Poisson
bracket with the constraints is weakly zero.

In the case of the BV solution the following gauge is fixed for the connection
\be
A_1 = A_2 = A \quad,\quad
A_3 = A + c_0 \exp ( -\int_0^\th  d\theta A ) \quad,\quad c_0 \ne 0
\quad, \label{25}
\ee
where $A$ is an arbitrary smooth function of $\theta$, such that $A \ne 0$
in $(a_k , b_k)$, where $0 < a_1 < b_1 < ... < a_n < b_n < 2\pi$. The 
intervals $(a_k , b_k)$ are called null patches, since the 
conection curvature vector $F\sp a = \e\sp{ij}F_{ij}\sp a$ is null there
($f_2 = f_3$). In that case the constraints give
\be
E_3 = E_2 \quad,\quad
E_1\sp{\prime} = (A_3 - A_2 ) E_2 \quad.\label{27}
\ee
This solution belongs to the degenerate metrics sector, since  
\be \det ||g\sp{ij}|| = E_1\sp 2 (E_2\sp 2 - E_3\sp 2 ) = 0 \quad.\label{28}\ee
In the flat patches $(b_k , a_{k+1})$, $A=0$ and $F =0$, so that the
solution is given by
\be
A_3 = c_k \quad,\quad E_1\sp{\prime} = A_3 E_2 \quad,\quad
 E_2 = 1/c_k \quad,\quad E_3 \ne E_2 \quad,
\label{281}
\ee
where $c_k = A_3 (b_k)$, and $E_3$ is an arbitrary function satisfying the 
boundary conditions. 
The last condition in (\ref{281}) insures that the
metric is non-degenerate, so that
the trace of a holonomy of a loop (Wilson loop) in a flat patch 
\be W = tr U = tr\, P \exp \left( \int_0^{2\pi} A_\f d\f \right)\label{wl}\ee 
is an integral of motion given by 
$2\cos (\pi c_k)$. Since the solution in the null patches (\ref{27}) depends
on the unconstrained canonical pair $(A_2, E_2 )$, it follows that one can
specify points in the part of the reduced configuration space by arbitrary 
many 
independent parameters $(c_1, \cdots , c_n)$, and hence one concludes that the
dimension of the rps cannot be finite.

This would be true provided that the null-patch solutions (\ref{27}) with 
different
functions $A$ are gauge inequivalent. However,
this is not the case, which follows from the fact that the function $A$ must 
satisfy
\be  \int_0^{2\pi} d\theta A = 0 \quad.\label{282}\ee
This means that BV solutions belong to the $a_1 = 0$ class, and hence they
are all gauge equivalent to $A=0$ solution, for which $c_1 = \cdots = c_n$.
Hence the quantities $c_k$ are not gauge invariant objects, or observables,
although they are integrals of motion. This follows from the fact that
$W_k = 2\cos(\pi c_k)$ are traces of holonomies, and these are not 
observables in the
Ashtekar formulation, although they can be integrals of motion in the gauges
where $F=0$ in some parts of the torus, like in the BV case. 

One can check that the BV gauge is dynamically consistent,
which ammounts to imposing the time preservation of the gauge-fixing
functions
\be F_1 = A_1 - A_2 = 0 \quad,\quad F_2 = (A_3 - A_1)\sp{\prime} +
A_1 (A_3 -A_1) = 0 \quad.\label{29}\ee
From $dF_1 /dt = 0$ and $dF_2 /dt = 0$, the equations of motion imply
\be -d\l\sp 1 /d\th + A_3 \l\sp 1 = f_2 (\l\sp 2 + E_1 \l\sp 3 ) \quad.
\label{210}\ee
By further requiring that the relations (\ref{27}) are preserved in time
one finds that  the Lagrange multipliers are fixed as
\be \l\sp 1 = 0 \quad,\quad \l\sp 2 = - E_1 \l\sp 3 = - C/2 (A_3 - A_1)\sp{-1}
 \quad,\label{212}\ee
where $C$ is an arbitrary constant. In the flat patches the dynamical 
consistency requires $\l_1 =0$, so that $\dot A_2 = \dot A_3 = 0$ and hence
the corresponding Wilson loop is independent of time. However, since the
Wilson loop is not an observable in the Ashtekar formulation, one can find a
gauge where it will  be time dependent. This can be seen from the following
derivation. By using a formula
\be  \exp (2\th_2 J_2 + 2\th_3 J_3 ) = {\rm ch}\sqrt{\th_2\sp 2 - \th_3\sp 2 } 
+ {{\rm sh}\sqrt{\th_2\sp 2 - \th_3\sp 2 }\over\sqrt{\th_2\sp 2 - \th_3\sp 2 }}
(2\th_2 J_2 + 2\th_3 J_3 )\, ,\label{hol} \ee
the Wilson loop (\ref{wl}) can be evaluated explicitely
\be  W = tr \exp (2\pi A_2 J_2 + 2\pi A_3 J_3 ) = 
2{\rm ch}\left(\pi\sqrt{A_2\sp 2 -A_3\sp 2}\right)\quad.\label{wl1}
\ee 
In a flat patch one then has
\be 
\dot W = -4\pi 
{{\rm sh}\left(\pi\sqrt{A_2\sp 2 - A_3\sp 2 }\right)
\over\sqrt{A_2\sp 2 - A_3\sp 2 }} 
\l\sp 1 A_2  A_3 \quad, \label{wl2}\ee
from which is clear that $W$ will be time 
dependent in gauges where $\l\sp 1 A_2 A_3 \ne 0$. An example of such a gauge
is
\be A_1 = 0 \quad,\quad  A_2 = \a (\theta) + d_k \quad ,\quad
 A_3 = \a (\theta) + h_k \quad,
\label{ng}\ee
where $d_k \ne h_k$ are constants, and $\a (\th)$ takes constant values in 
flat patches, so that $f_2 = f_3 =0$, while in  null patches 
$\a\sp{\prime} (\th) \ne 0$ and hence $f_2 = f_3 \ne 0$.
The gauge (\ref{ng}) is an alternative realisation of the null and flat patch
initial data, and the dynamical consistency requires that $\l_1 = -1$. In a
flat patch one then  has
\be \dot A_1 = 0 \quad,\quad\dot A_2 =\l\sp 1 A_3 \quad,\quad \dot 
A_3 = - \l\sp 1 A_2 \quad,
\ee 
so that 
\be A_2 = \a_k \cos (\l\sp 1 t + \varphi_k ) \quad,\quad 
A_3 = \a_k \sin (\l\sp 1 t + \varphi_k ) \label{tda}\ee
where $\a + d_k = \a_k \cos\varphi_k$ and $\a + h_k = \a_k \sin\varphi_k$. 
Hence the Wilson 
loop (\ref{wl}) is not an observable, since it is possible to find gauges 
where $\dot W \ne 0$.

\section{Residual gauge transformations}

Our arguments from the previous section imply that the BV solution with
arbitrary $c_k$ coefficients is gauge equivalent to $c_1 = ... = c_n$
solution. This means that the gauge (\ref{29}) has an infinite-dimensional 
residual 
gauge invariance. One can find this residual gauge invariance from the
requirements $\d F_1 = \d F_2 =0$. They give
\be -d\e\sp 1 /d\th + A_3 \e\sp 1 = f_2 (\e\sp 2 + E_1 \e\sp 3 )\quad.
\label{rt}\ee
Equation (\ref{rt}) implies that the gauge parameters $\e\sp 2$ and 
$\e\sp 3$  are 
not fixed, so that the 
corresponding gauge transformations will preserve the gauge (\ref{29}). 
This is a direct consequence
of the fact that in the BV gauge the constraints $G_2$ and $G_3$ become
linearly dependent. Note that in a non-null gauge ($ f_2 \ne \pm f_3$),
where the constraints are independent, 
the residual transformations for static solutions are given by 
\bea
\d A_1 &=& -{d\e\sp 1 \over d\th}  = 0 \quad,
\label{2131}\\
\d A_2 &=& -A_3\e\sp 1 + f_2 \e\sp 2 + E_1 f_3 \e\sp 3 = 0
\quad, \label{2132}\\
\d A_3 &=& A_2 \e\sp 1 - f_3 \e\sp 2  - E_1 f_2 \e\sp 3  = 0
\quad. \label{2133}
\eea
These equations completely fix the $\e$'s, and consequently there is no
residual gauge invarince, which corresponds to the fact that in such a gauge
the solution contains only the physical degrees of freedom.

This analysis implies that the Wilson loop (\ref{wl}) will not be invariant
under the residual gauge transformations in flat patches. This can be seen from
the expression (\ref{wl1}) since
\be 
\d W = -4\pi 
{{\rm sh}\left(\pi\sqrt{A_2\sp 2 - A_3\sp 2 }\right)
\over\sqrt{A_2\sp 2 - A_3\sp 2 }} 
\e\sp 1 A_2  A_3 \quad. \label{wl3}\ee
In the BV gauge this variation is zero, since $A_2 = 0$. However, the residual
transformations (\ref{rt}) allow one to choose a gauge in a flat patch such 
that $A_2 \ne 0$, and hence the variation (\ref{wl3}) will be non-zero.
Really, the infinitesimal gauge transformations (\ref{rt})
imply that the BV configuration $A_1 =A_2 = 0$ and $A_3 = c_k$ transforms into
\be A_1 = A_2 = -c_k \e\sp 1 (\th) \quad,\quad A_3 = c_k \quad,\label{ic}\ee 
for which the variation
(\ref{wl3}) is non-zero, and it is of order $(\e\sp 1)\sp 2$.
This means that the variation (\ref{wl3}) will be non-zero at the second
order in $\e\sp 1$ if one starts from the $A_2 = 0$ configuration.
Note that the  configuration (\ref{ic}) corresponds to  the lowest order terms
in the $\e\sp 1$ expansion of the configuration
\be A_1 = A_2 = - A_3 = {1\over \th - \th_k } \quad,\label{bc}\ee
where $\th_k$ is a constant. This follows from the fact that the
equations
\be f_2 = A\sp{\prime} - A A_3 = 0 \quad,\quad 
f_3 = A_3\sp{\prime} - A\sp 2 = 0 \quad,\ee
have (\ref{bc}) as the unique solution for $A \ne 0$. In the configuration 
(\ref{bc}) one finds that $W (A) = 2$, and hence $c_1 = .... = c_n = 0$. 
Therefore $W(A) \ne W(\tilde A)$, 
where $A$ and $\tilde A$ are related by a gauge transformation, which confirms
our result from the previous section that the Wilson loop is not an
observable, and also shows that the configuration with arbitrary $c_k$'s is
gauge equivalent to $c_k = 0$ configuration.

Therefore the gauge inequivalent BV solutions are 
labeled with only one parameter $e_1$, which is canonicaly conjugate to the
gauge invariant parameter $a_1$. The corresponding phase space is
labeled by $(a_1, e_1)$, which is a subspace of the total rps given locally by
the four-dimensional symplectic manifold ${\bf R}\sp 4$ \cite{mm}.

\section{Conclussions}

We have  demonstrated
that the number of physical dof for (2+1)-dimensional Ashtekar gravity on 
a torus is finite,  which confirms the correctness of the ansatz of \cite{mm}. 
It would be an interesting problem to show that the number of physical dof
is finite for a higher-genus $\S$. 

It is important to realize that
an integral of motion does not have to be an observable, and this is the
reason why Barbero and Varadarayan have made a wrong conclussion about the
dimensionality of the rps. 
The more familiar examples of this situation are the
static solutions of 2d dilaton gravities, which include the spherically
symmetric GR solutions. In that case the 
components of the metric are independent of 
time, and hence they are integrals of motion. However, the metric is obviously
not an observable. The observable is the ADM mass, which labels the gauge 
inequivalent solutions. In the BV case, the role of the ADM mass is played by 
$a_1$. Also, in analogy to the gauge (\ref{tda}), one can find gauges
in the 2d dilaton gravity case where the metric is time dependent.

From the point of view of the full Ashtekar theory, the
Wilson loop $W$ can be an observable only if $F = 0$ 
and ${\rm det} E \ne 0$ conditions can be preserved in time independently
of the gauge, or equivalently, of the 
value of the Lagrange multipliers. However, although the $F=0$ condition can be
preserved in time for every initial flat configuration, 
the metric non-degeneracy condition ${\rm det}E \ne 0$
can not be preserved in time for every inital configuration, 
which explains why in the case of the
gauge (\ref{ng}) the Wilson loop becomes time dependent in a flat 
patch.

Note that the BV solution 
can be understood from the general theory of gauge fixing
(see \cite{mm2} and references there) in the following way. 
For a dynamical system with first-class irreducible 
constraints $G_\a (p,q)$, the rps is obtained by choosing the gauge fixing 
conditions $\chi\sp\a (p,q) = 0$\footnote{In the case of a reparametrization
invariant system, this condition must be generalized to 
$\chi\sp\a (p,q) = \d_1\sp \a f(t) $, where $t$ is the evolution parameter
\cite{mm2}. This amounts to choosing the time variable in the system.
The gauge choice (\ref{25}) should be modified accordingly, but this does not
affect the subsequent analysis.} 
such that
\be \{ \chi\sp\a , \chi\sp\b \} = 0 \quad,\label{31}\ee
and the Faddeev-Popov determinat
\be \D = \det \{ G_\a , \chi\sp\b  \}|_{\chi = 0} \quad,\label{32}\ee
must be different from zero. In the infinite-dimensional case, $\D$ has to
be carefully defined, because of the presence of the trivial zero modes, which
should be omitted. Therefore the condition of non-zero $\D$ in the infinite
dimensional case means that the operator $\hat \D$, defined
by $\{ G(x_1) , \chi (x_2) \}|_{\chi = 0}$, must not have non-trivial zero
eigenvalues. This translates into examining the solutions of the equation
($\a = \{ i,x \}$)
\be {\hat \D}\sp{i}\sb{j} \e\sp j (x) = 
\int dx_1 \e\sp j (x_1) \{ G_j (x_1) , \chi\sp i (x) \}|_{\c =0} = 0 \quad.\ee
But this is precisely the condition for finding the residual gauge invariances.
Finding precise criteria for a non-trivial solution depends on a concrete
theory, but
it is clear that a solution for which one or more of $\e\sp i (x)$
are unrestricted is non-trivial. Such solutions appear in the
gauge (\ref{25}), which means that a gauge
is chosen such that the equations $G_i |_{\c =0} = 0$ are linearly dependent.
In other words, when solving the constraints in such gauges, one uses only
a part of the constraints, and that is the reason why one obtains the 
solutions with more dof than the number of physical dof.

\end{document}